\pdfoutput=1

\documentclass[prresearch,twocolumn, floatfix, superscriptaddress]{revtex4-2}

\usepackage{graphicx}
\usepackage{dcolumn}
\usepackage{bm}
\usepackage{tabularx}
\usepackage{booktabs} 

\begin{document}

\title{Deterministic Mapping of Topological Phases via Autoregressive Exogenous Neural Networks}

\author{Graciana Puentes}
 \affiliation{Departamento de F\'isica, Facultad de Ciencias Exactas y Naturales, Universidad de Buenos Aires, Buenos Aires, Argentina}
 \affiliation{CONICET-Universidad de Buenos Aires, Instituto de Física de Buenos Aires (IFIBA), Buenos Aires, Argentina}

\date{\today}

\begin{abstract} 
We report a comparative analysis of three dynamic neural network (NN) architectures—NAR, NARX, and NIO—to evaluate their efficiency in estimating the critical-measurement-strength parameter ($c_{\mathrm{crit}}$) characterizing topological phase transitions in geometric phases induced by weak measurements. Our results demonstrate that the NARX architecture achieves superior predictive fidelity, reaching a Mean Squared Error (MSE) of $10^{-27}$ - the limit of numerical precision- at an optimal delay of $d=1$. This exceptional performance implies the identification of a perfect functional identity, suggesting that the relationship between winding numbers $W$ and $c_{\mathrm{crit}}$ is mathematically deterministic. We observe a  \textquotedblleft complexity paradox" where the NARX model's accuracy collapses at higher delays ($d=4$), a phase-sensitivity that confirms the model captures a high-precision dynamic mapping rather than a trivial pattern. While the NAR model remains robust for local-trend capture, the NIO architecture fails to accurately resolve the phase transition despite increased neuronal capacity. These findings underscore that both autoregressive feedback and immediate exogenous context are essential for the exact characterization of topological phases, establishing NARX as a robust framework for deriving governing laws in complex quantum systems, where analytical solutions remain elusive.\end{abstract}

\maketitle


\section{Introduction}

The emergence of global quantum phases represents a cornerstone of modern physics, providing the bedrock for our understanding of gauge-invariant observables. As first elucidated by Sir Michael Berry, when a quantum system undergoes an adiabatic cycle, it accumulates a geometric phase that is fundamentally independent of the dynamical evolution, depending instead on the global topology of the path traced in parameter space \cite{1,2, 3, Berrygraphene, Berrytopoinsul, Berryelectronic, Berrychemestry, GeomPhase1, GeomPhase2, GeomPhase3}. These phases are not merely theoretical constructs; they dictate the conductivities of graphene, the robust edge states of topological insulators, the branching ratios of molecular chemistry, and the fractional statistics of anyonic quasiparticles \cite{4, 5, 6, Hall1, TopoInsu1, TopoInsu2, ZakDemler, ZakLonghi, PuentesJOSAB, PuentesCrystal, PuentesEntropy, PuentesQuantumRep, PuentesFrontPhys, BerryStack}.

While the Pancharatnam phase—arising from sequences of projective measurements—has been extensively characterized in the limit of the quantum Zeno effect \cite{Pancharatnam, Zeno}, the behavior of geometric phases induced by weak measurements remains a frontier of quantum control. Weak measurements offer a unique window into the continuous tracking of quantum states with minimal back-action, enabling the observation of trajectories that would otherwise collapse under strong projection \cite{WM1, WM2, WM3}.

In this work, we investigate a novel class of geometric phases induced by weak measurements, specifically focusing on the critical transitions governed by the winding number $W$ of the polar angle $\phi$. Unlike previous approaches focusing on singular trajectories \cite{2}, we address the fundamental question of whether the \textquotedblleft stochastic" jumps at the critical-measurement-strength parameter associated with topological transitions are truly independent, or if they are governed by a hidden, deterministic law shared across different topological sectors.

To resolve this, we leverage the predictive power of deep learning architectures—specifically Nonlinear Autoregressive with Exogenous Inputs (NARX), Nonlinear Autoregressive (NAR), and Nonlinear Input-Output (NIO) neural networks \cite{RNN1, RNN2, NARX1, NARX2, NARX3, NAR1, NAR2, CNN1, CNN2}. While traditionally employed for time-series forecasting in fields ranging from robotics to finance \cite{Omolayo}, we repurpose these architectures as \textquotedblleft numerical probes" to quantify the informational entropy of topological transitions.

We present a comparative analysis that establishes a rigorous performance hierarchy between these models. Most significantly, we demonstrate that the NARX architecture, when supplied with exogenous data from lower winding numbers ($W=1…4$), can predict the critical-parameter transition of $W=5$ with a precision of $10^{-27}$ —effectively reaching the limit of numerical representation. This near-perfect convergence serves as a numerical existence proof for a deterministic functional identity: it reveals that the information governing a phase transition in a specific topological sector is non-locally encoded within the global context of its predecessors. Our results suggest that neural networks can act as highly accurate deterministic mapper unearthing exact analytical solutions hidden within complex quantum measurement data and providing a new framework for characterizing the memory and context-dependence of topological quantum matter.

\begin{figure*}[t]
\centering
\includegraphics[width=0.9\textwidth]{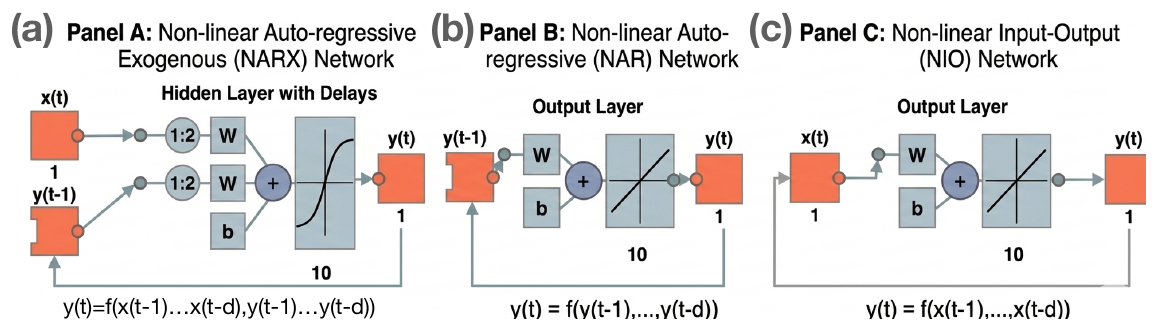}
\caption{(a) Panel A: Non-Linear Auto-regresssive Exogenous (NARX) Neural Network governed by the generic mapping $f(x(t-1),...,x(t-d), y(t-1),...,y(t-d))$, (b) Panel B: Non-Linear Autoregressive Neural Network (NAR) governed by the generic mapping $f(y(t-1),...,y(t-d))$, (c) Panel C: Non-Linear Input-Output (NIO) Neural Network governed by the generic mapping $f(x(t-1),...,x(t-d))$. }
\end{figure*}

\section{Mathematical model}

Building upon recent advances in topological geometric phases induced by weak measurements \cite{2}, we established an analytical and numerical framework for a novel class of geometric phases induced by weak measurements over distinct topological sectors, characterized by the winding number $W$ \cite{PuentesFrontPhys}. While weak measurement protocols are recognized for their ability to provide continuous state evolution with minimal back-action \cite{WM1,WM2,WM3}, our approach distinguished itself from existing models—such as the exponential approximation in \cite{2}—by explicitly considering the dependence of the geometric phase on the winding number $W$ of the polar angle $\phi$. In our study, the system was initialised in the equatorial state $|\psi_0 \rangle = \frac{| \uparrow  \rangle + | \downarrow \rangle}{\sqrt{2}}$  and subjected to a sequence of weak measurements in $\phi$. By parameterizing the rotation as $\alpha =\epsilon N/2$, with $N$ the dimension of the Hilbert space, it was demonstrated that the trajectory closes at discrete values $\alpha=k\phi$, where the winding number is defined as $W=k$. A critical finding of that work was that different winding numbers $W$ yield unpredictable critical-measurement-strength parameters ($c_{\mathrm{crit}}$). At these points, the geometric phase exhibits discrete $|\pi|$ jumps, signalling non-equivalent topological phase transitions. These results provide the necessary topological foundation for neural-network-assisted critical-measurement-strength parameter estimation,  investigated in the present work.

The measurement sequence required to accumulate the intended geometric phase $X$ can be mathematically described by a complete set of POVMs (Positive Operator-Valued Measures),
 implemented via the Kraus operators  $\mathcal{M}_{k}^{(r_{k})}=M_{\eta_{k}}({\bf{n_{k}}},r_{k})$, $|\psi \rangle \rightarrow \mathcal{M}_{k}^{(r_{k})} |\psi \rangle$, as described in \cite{2}. 
 Such POVM can be implemented by introducing a detector consisting of a second qubit whose Hilbert space is spanned by the set $r=\{ |+ \rangle, |- \rangle\}$. We consider the generic initial state of the system of the form $|\psi_0 \rangle = a |\uparrow \rangle +b|\downarrow \rangle$, and assume the detector is in the initial state $|+ \rangle$ and that the initial state of the system plus detector is separable, of the form $|\psi_{\mathrm{sep}}\rangle = |\psi_0 \rangle \otimes |+ \rangle$. The measurement coupling $\lambda(t)$ is then switched on for a finite time $t \in [0,T]$, to obtain the entangled state:

\begin{equation}
|\psi_{\mathrm{ent}}\rangle = M_{\eta}({\bf{n}},+) |\psi_0 \rangle +  M_{\eta}({\bf{n}},-) |\psi_0 \rangle,
\end{equation}

here the measurement strength is $\eta \propto \sin^2(g)$, with $g=\int_0^{T} \lambda(t)dt$. The POVMs describing the measurement process are defined by the Kraus operators \cite{2}:

\begin{equation}
 M_{\eta}(\hat{z},+)= 
\left(\begin{array}{cc}
 1 &  0 \\ 
 0 &  \sqrt{1-\eta}
   \end{array} \right) \hspace{0.5cm}   M_{\eta}(\hat{z},-)= \left(\begin{array}{cc}
 1 &  0\\
 0 &  \sqrt{\eta}
  \end{array}\right)
  ,
  \end{equation}
  
 corresponding to a measurement orientation along the $z$-axis ${\bf n}=\hat{z}$. 
 
 
 
 An analytic expression for the geometric phase $X$ can be derived in the quasi-continuous measurement limit $N \rightarrow \infty$. $X$ is extracted from the quasi-continuous trajectory postselecting all outcomes $r_{k} = |+\rangle$ as described in PNAS2020 \cite{2}. In our case, we consider the initial state to be of the form $|\psi_0 \rangle = \frac{| \uparrow  \rangle + | \downarrow \rangle}{\sqrt{2}}$, this is equivalent to setting the initial parameters $(\theta_0=\pi/2, \phi_0=0)$. We then sequentially rotate the measurement apparatus, in order to increment the angle $\phi$   by a fixed amount $\epsilon=2 \pi/N$. This parameterization ensures that the trajectory is closed and the geometric phase well defined, resulting in an analytic expression for the acquired geometric phase \cite{PuentesFrontPhys}:

 \begin{equation}
 X= e^{-i \alpha -c}[\cosh(\tau) + c \sinh(\tau)/\tau \textbf{]},
 \end{equation}

 where $\tau=\sqrt{c^2 - \alpha^2}$ and $\alpha= \epsilon N/2$. We recover the analytic result reported in  \cite{2}, by setting $z=c$, and $\alpha=\pi$, as expected for $\theta=\pi/2$.  \\

We note that our case is complementary to the one considered in PNAS2020 \cite{2}, where the authors analyzed the dependence of the geometric phase on the azymmuthal angle $\theta$ for a single winding of the trajectory with the polar angle $\phi$. Here we analyze the dependence of the geometric phase on the winding of the trajectory by the steering sequence of weak measurements of increasing angle $\phi$, obtained by increasing the  winding number $W$, while fixing $\theta=\pi/2$. In Ref.\cite{PuentesFrontPhys} we stated that the exact critical-measurement-strength parameter $c_{\mathrm{crit}}$ for different winding numbers $W=1,2, 3,...,M$ (with $M \in Z^+$) was not predictable to our knowledge. One of the main findings of this work is that we demonstrate that a mathematically deterministic relationship between $W$ and $c_{\mathrm{crit}}$ indeed exists, utilising dynamic NARX neural network.   \\

\section{Data Representation and Topological Sequences}

\begin{figure}[t]
\centering
\includegraphics[width=1\linewidth]{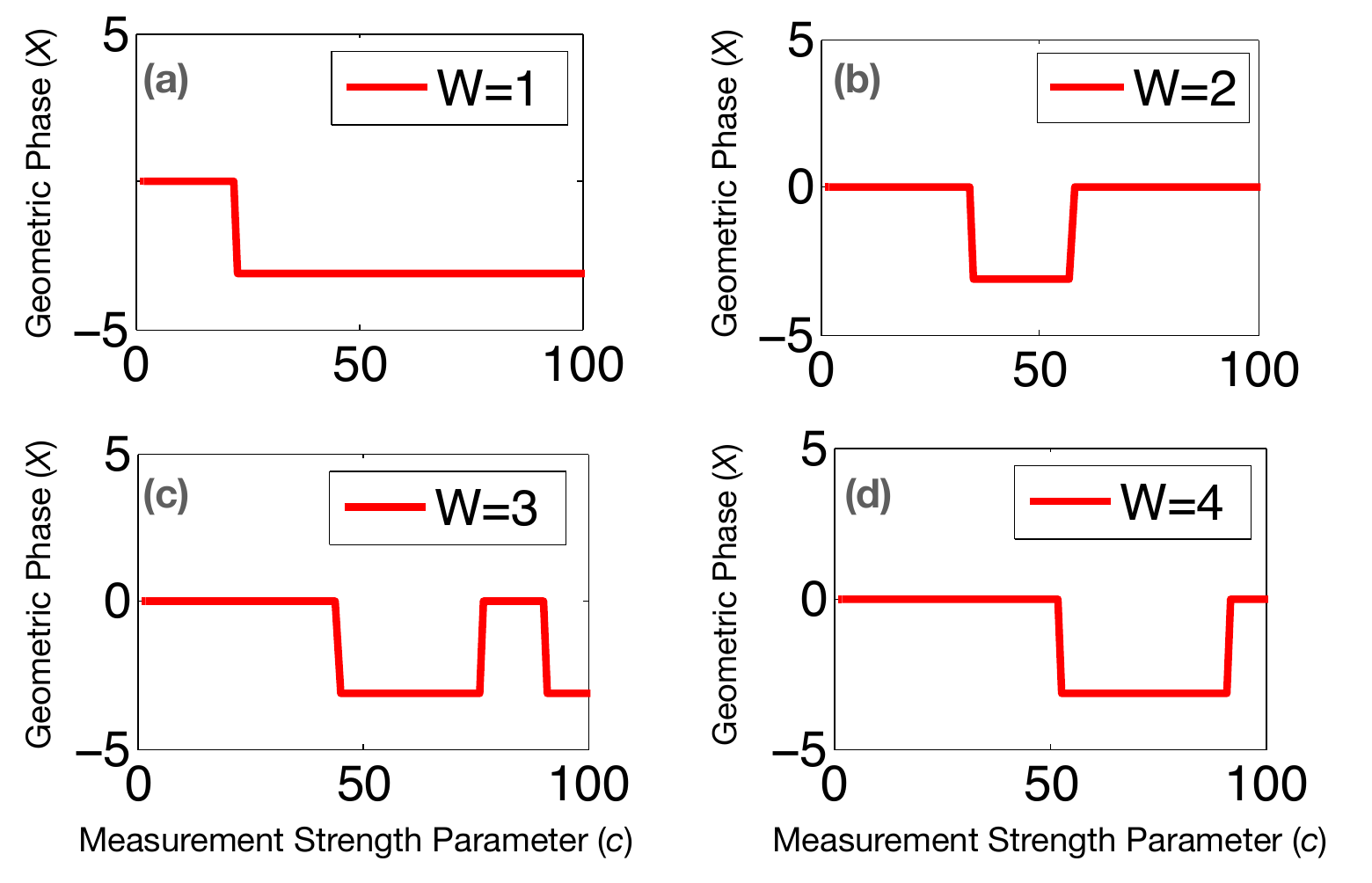}
\caption{\label{fig:frog} Exogenous sequences ($x$) used to train, test, and validate the NARX and NIO neural networks. The sequences consist of $|\pi|$ discrete jumps in the geometric phase ($X$) vs. measurement-strength parameter  ($c$) for increased winding number $W=1,2,3,4$:  (a) $(W=1)$ $c_{\mathrm{crit}}=21$, (b) $(W=2)$ $c_{\mathrm{crit}}=34$ and $c_{\mathrm{crit}}=57$, (c) $( W=3)$ $c_{\mathrm{crit}}=44$ and $c_{\mathrm{crit}}=76$, (d) ($W=4$) $c_{\mathrm{crit}}=52$ and $c_{\mathrm{crit}}=91$ and  As we restrict our analysis to trajectories on the equator, the acquired geometric phase ($X$) can only take values of 0 or $| \pi|$ (modulo  $2 \pi$).The endogenous (target) sequence ($y$) used in all the three networks corresponds to the acquired geometric phase ($X$) vs. measurement-strength parameter ($c$), for a winding number $W=5$.}
\end{figure}

The identification of topological phases requires the resolution of invariants that remain robust under continuous deformations of the system’s Hamiltonian. In our framework, the geometric phase $X$ serves as the order parameter, while the measurement-strength parameter $c$ acts as the driving coordinate. Any sequence of observables indexed by a strictly monotonic parameter—whether temporal, spatial, or, as in this study, a coupling strength—defines a trajectory on a dynamical manifold. The structural correspondence between these sequences and classical time-series arises from the autocorrelation of neighbouring states along the parameter’s evolution, where the state at $c_{i}$  is functionally constrained by the history of the trajectory $\{c_0,...,c_{i-1}\}$.  
This parameter-space progression allows us to deploy dynamic neural network architectures, such as NARX and LSTM, to model the non-linear dependencies and non-analyticities inherent in topological jumps. By abstracting the coupling parameter $c$ as an evolutionary axis analogous to $t$, the network learns to map the underlying \textquotedblleft equations of motion" of the geometric phase, effectively identifying the critical thresholds $c_{\mathrm{crit}}$, where the system undergoes a discrete topological jump.

We characterize the system using a set of training sequences defined by the acquired geometric phase $X(c)$ across varying topological sectors, denoted by the winding number $W$. Following the formalism described in Ref. \cite{PuentesFrontPhys}, our analysis is restricted to trajectories constrained to the equator of the Bloch sphere ($\theta=\pi/2$). In this configuration, the acquired phase over a closed trajectory $C$ along the equator is a discrete topological observable:
\begin{equation}
X= \oint_C A.d\textbf{R},
\end{equation}
where $A$ is the Berry connection. $X$ can only take discrete values equal to $0$($\pi$) for an open(closed) trajectories. The transition between these values $(0,\pi)$ defines the critical measurement strength $c_{\mathrm{crit}}$. 

As illustrated in Fig. 2, the training set comprises exogenous sequences for $W \in {1,2,3,4}$, each displaying characteristic phase-flip signatures:
\begin{itemize}

\item{(a) } $W=1$: A single transition at $c_{\mathrm{crit}} =21.$

\item{(b)} $W=2$: Bifurcated transitions at $c_{\mathrm{crit}}=34$ and $57$.

\item{(c)} $W=3$: Transitions at $c_{\mathrm{crit}}=44$ and $76$.

\item{(d)} $W=4$: Critical points at $c_{\mathrm{crit}}=52$ and $91.$
\end{itemize}

These discrete jumps—representing the system's \textquotedblleft topological memory"—provide the high-frequency features necessary to test the network’s capacity for discontinuity resolution. The target sequence for all predictive models is the $W=5$ sector, allowing us to evaluate the network's ability to generalize the global laws governing topological transitions from lower-order training data.

\section{Dynamic Neural Network Architecture}

To investigate the deterministic dependencies within topological phase transitions, we employ a class of dynamic neural networks characterized by temporal state-space embeddings. The core architecture consists of a multi-layer perceptron (MLP) framework integrated with tapped delay lines (TDL), enabling the network to map the history of the system onto its future evolution. The hidden layer utilizes nonlinear transcendental activation functions—specifically the hyperbolic tangent ($\tanh$)—to project input features into a high-dimensional manifold, while the output layer remains linear to preserve the continuous range of the geometric phase $X$. The network dynamics are governed by the generic mapping:

\begin{equation}
y(t)=f[x(t-1,…,t-d),y(t-1,…,t-d)]+\epsilon(t),
\end{equation}

where $d$ denotes the embedding dimension (delay depth) and $\epsilon(t)$ represents the residual error. During the training phase, the architecture is implemented in a series-parallel (open-loop) configuration. In this regime, the ground-truth historical values of the geometric phase are supplied to the feedback loops, effectively decoupling the temporal recursions to provide a stable gradient surface for the optimization of the synaptic weights. For predictive inference, the network is transitioned into a parallel (closed-loop) mode; here, the network becomes a self-sustained dynamical system where predicted outputs are recursively fed back as inputs. This transition allows the model to simulate the autonomous evolution of the topological state, testing the robustness of the learned mapping against error propagation.

In our implementation, the sequence of the geometric phase $X$ vs. measurement-strength parameter $c$ for $W=5$ is defined as the target manifold $y(t)$. The exogenous input $x(t)$ is represented by a  vector of the phases for $W=1,2,3,4$. We systematically vary the neuronal density and the delay depth $d$ to explore the transition from local trend-following to global functional identity.

\subsection{Nonlinear Autoregressive with Exogenous Inputs (NARX)}

The NARX architecture treats the winding number $W=5$ as the endogenous target $y(t)$, while simultaneously consuming the sequences $W \in \{1,2,3,4\}$ as exogenous drivers $x(t)$. Mathematically, it assumes the system is non-autonomously driven:

\begin{equation}
y(t)=f(x(t-1),…,x(t-d),y(t-1),…,y(t-d)).
\end{equation}

This model is designed to test the hypothesis of topological entanglement, where the phase transition at higher winding numbers is functionally locked to the states of lower-order sectors. The NARX architecture is depicted in Fig, 1 (a), Panel A. 

\subsection{Nonlinear Autoregressive (NAR)}

The NAR model serves as our control for intrinsic predictability. By discarding exogenous inputs, it attempts to map the future of the geometric phase solely from its own local history:

\begin{equation}
y(t)=f(y(t-1),…,y(t-d)).
\end{equation}

This architecture quantifies the self-correlation of the topological jump and identifies the limit of information contained within a single topological sector. The NAR architecture is depicted in Fig. 1 (b), Panel B. 

\subsection{Nonlinear Input-Output (NIO)}

The NIO network functions as a purely stochastic transducer, mapping external drivers to a target output without a feedback mechanism:

\begin{equation}
y(t)=f(x(t-1),…,x(t-d)).
\end{equation}

By omitting the autoregressive term, the NIO model tests whether the topological transition can be modeled as a memoryless (Markovian) response to external measurement strength, or if the "memory" of the phase transition is a requisite for convergence. The NIO architecture is depicted in Fig. 1 (c), Panel C. 

\begin{figure}[h]
\centering
\includegraphics[width=1\linewidth]{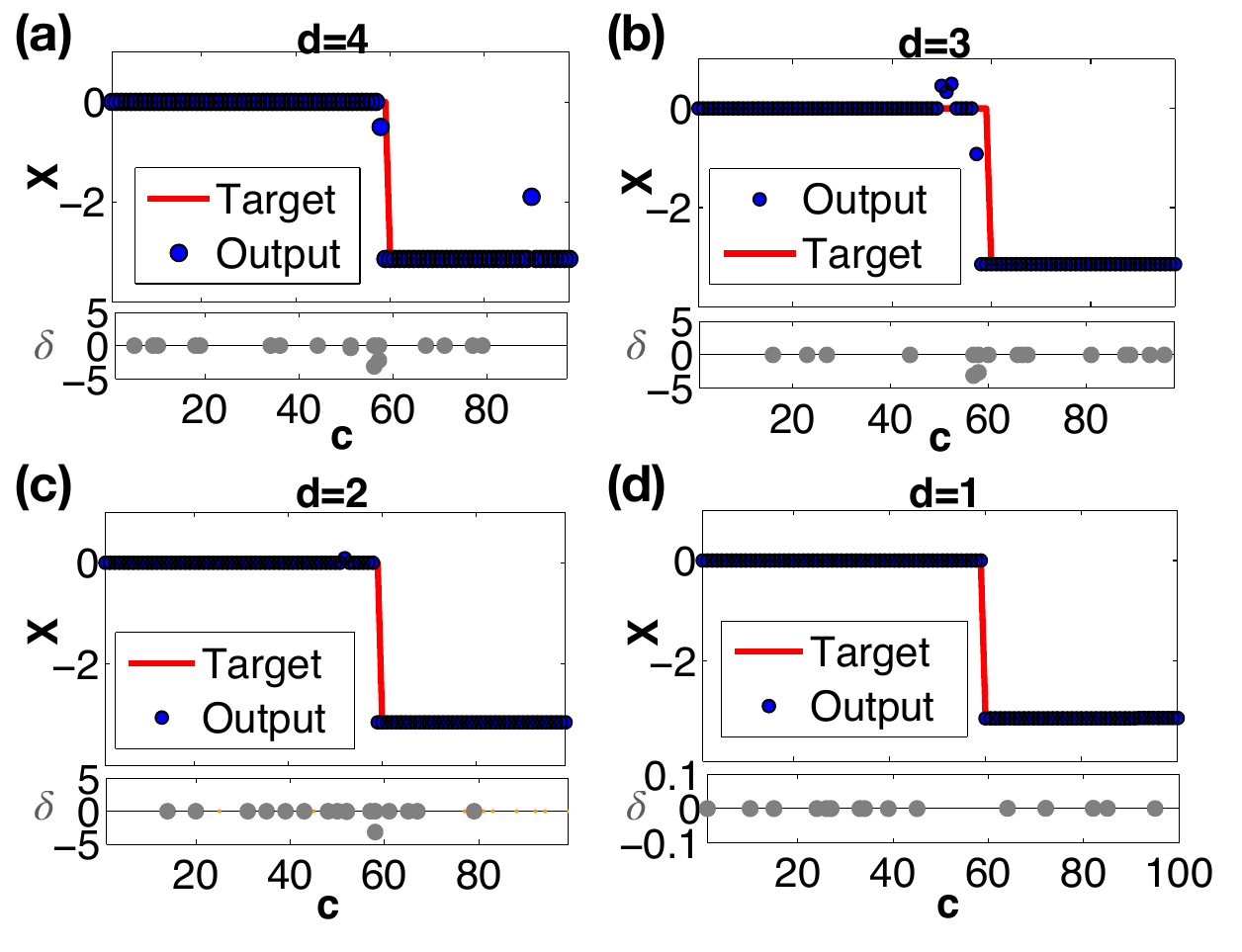}
\caption{\label{fig:frog} NARX neural network response as quantified by output (blue dots), target (red line) and absolute error ($\delta$, gray dots): (a) $d=4$ past values, (b) $d=3$ past values, (c) $d=2$ past values and (d) $d=1$ past values, $\delta$ is exactly equal to zero on any scale.  }
\end{figure}

\begin{figure}[h]
\includegraphics[width=1\linewidth]{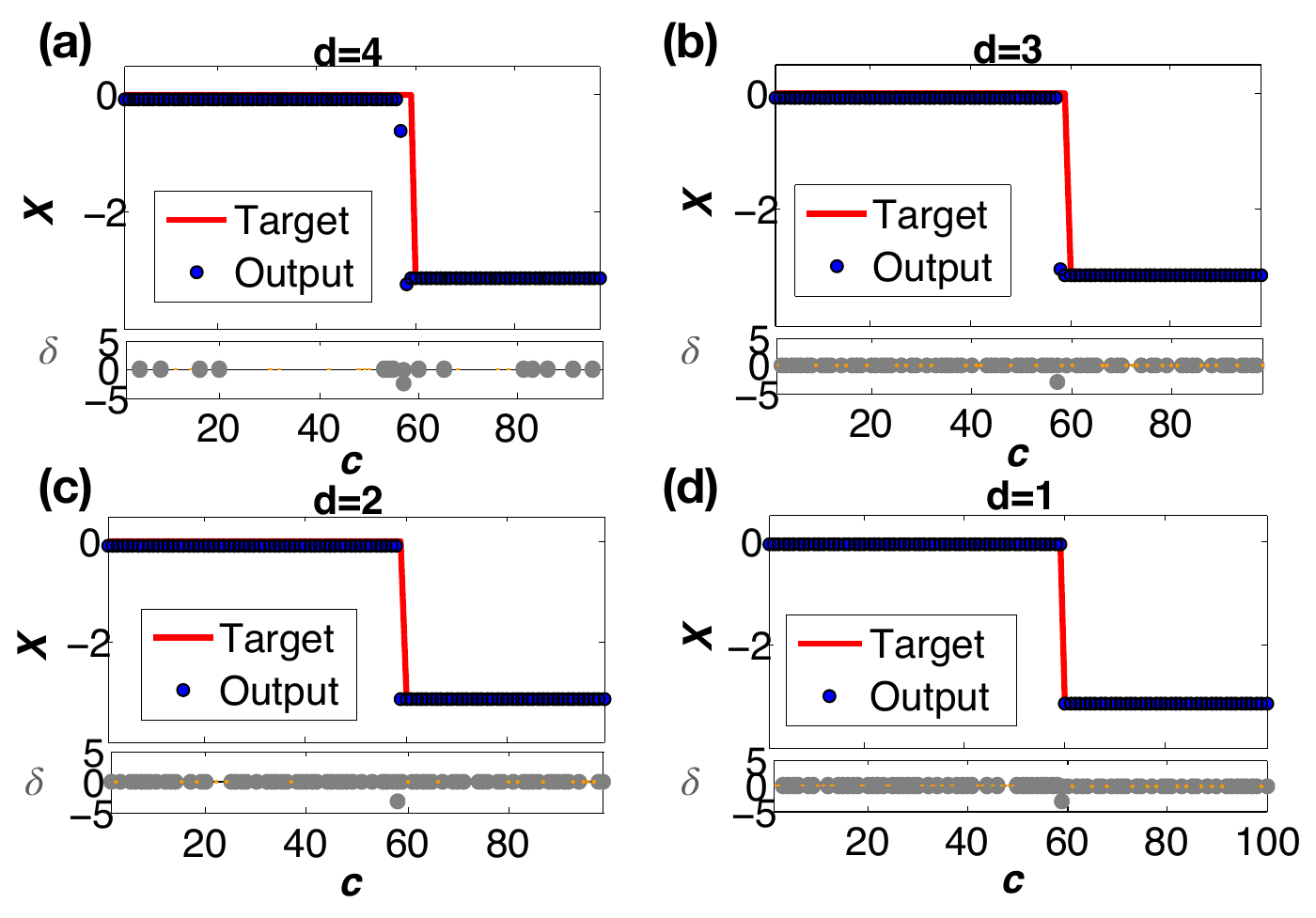}
\caption{\label{fig:frog}  NAR neural network response as quantified by output (blue dots), target (red line) and absolute error ($\delta$, gray dots): (a) $d=4$ past values, (b) $d=3$ past values, (c) $d=2$ past values and (d) $d=1$ past values. }
\end{figure}

\begin{figure}[hb]
\includegraphics[width=1\linewidth]{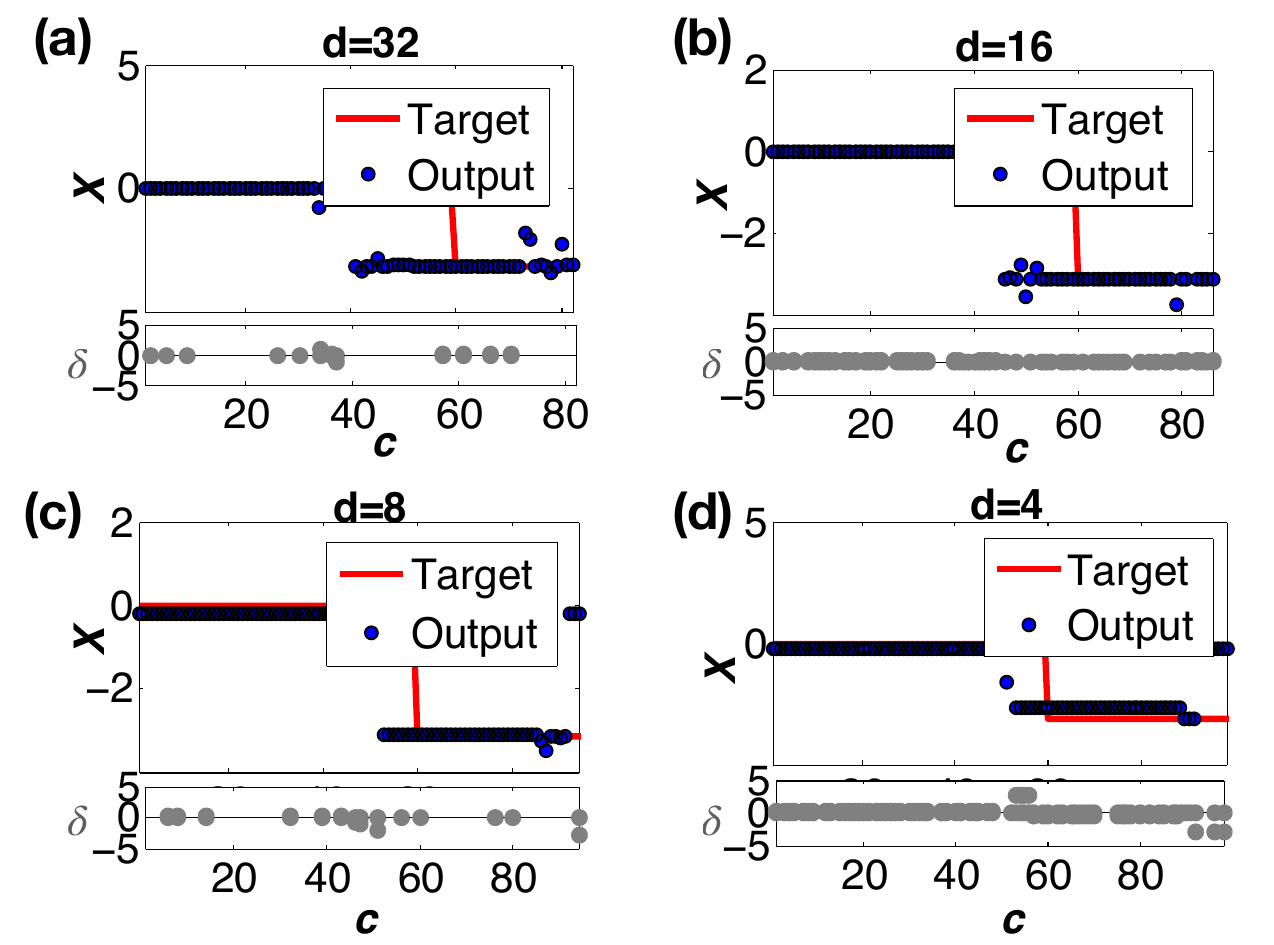}
\caption{\label{fig:frog} NIO neural network response as quantified by output (blue dots), target (red line) and absolute error ($\delta$, gray dots): (a) $d=32$ past values, (b) $d=16$ past values, (c) $d=8$ past values and (d) $d=4$ past values. }
\end{figure}

\section{Numerical Results and Comparative Analysis}

In this section we present the numerical results obtained using the dynamic neural networks (NNs) described in the previous sections. We quantify the NN predictive power by analysing  their response, their auto-correlation error and their performance, as described below. 

\subsection{Topological Response and Error Manifolds}

The performance of the dynamic architectures is first assessed through a direct comparison between the predicted Output and the Target manifold ($X(c)$). In the context of quantum phase transitions, a high-fidelity response is defined by the network's ability to localize the non-analytic \textquotedblleft jump" at $c_{\mathrm{crit}}$ without introducing artificial phase-lag or damping the discrete amplitude of the signal ($|\pi|$). We define the residual error manifold as $\delta(c)=X_{\mathrm{output}}(c)-X_{\mathrm{target}}(c)$. The statistical distribution of $\delta(c)$ serves as a primary diagnostic for the network's informational completeness:

\begin{itemize}

\item{Optimal Convergence}: A model that has successfully captured the underlying deterministic mapping will yield residuals characterized by Gaussian white noise centered at zero. This indicates that the network has internalized the governing equations of the geometric phase, leaving only irreducible numerical fluctuations.

\item{Systematic Divergence:} Conversely, the presence of structured patterns or periodic trends in $\delta (c)$ signifies a failure to resolve specific temporal dynamics or an inability to bridge the \textquotedblleft memory gap" between W sectors.

\item{Transition Volatility: }Localized spikes in the error manifold are particularly significant at the critical points $c_{\mathrm{crit}}$. These spikes quantify the network’s sensitivity to high-volatility topological events and regime shifts, providing a measure of the effective \textquotedblleft spectral resolution" of the NN architecture.

\end{itemize}

By analyzing the distribution of these residuals, we can distinguish between models that merely interpolate the local trend (NAR) and those that achieve a global functional identity through exogenous entanglement (NARX).

\subsection{Statistical Independence and Residual Whiteness}

The completeness of the neural network’s mapping is rigorously tested through the autocorrelation function (ACF) of the residuals. To confirm that the architecture has internalized the full dynamical evolution of the geometric phase, the residual sequence $\delta(t)$ must be indistinguishable from a stochastic white-noise process. We define the temporal autocorrelation $R (k)$ at lag $k$ as:
\begin{equation}
R(k)= \frac{\sum_{t=1}^{N-k} (\delta(t) -\bar{\delta})(\delta(t+k)-\bar{\delta})}{\sum_{t=1}^{N} (\delta(t) -\bar{\delta})^2},
\end{equation}
where  $\bar{\delta}$ is the mean residual and $N$ is the sequence length. In a theoretically optimal mapping, the ACF should satisfy the condition $R(k)=\delta_{k,0}$ (where $\delta_{k,0}$  is the Kronecker delta). This manifests as a solitary unitary peak at $k=0$, signifying that the error at any coupling strength $c$ is statistically independent of its history. Such a result provides a numerical guarantee that the NN has successfully filtered all systematic dependencies—whether local (autoregressive) or global (exogenous)—from the topological data. Diagnostic interpretation of ACF: 

\begin{itemize}

\item{Whiteness Limit:} When the ACF values for $k>0$ remain within the $95\%$ confidence bounds, typically defined as $\pm 2/\sqrt{N}$, the residuals are considered "white." This confirms that the model has reached the informational limit of the data, leaving only irreducible numerical or quantum fluctuations.

\item{Systematic Residual Memory: }Statistically significant spikes outside these bounds indicate "bad" autocorrelation, where the model fails to resolve the inherent structure of the phase transition. For instance, persistent correlations at low lags ($k=1$) suggest an insufficient temporal embedding depth $d$, implying that the network's "memory" is too short to bridge the non-analyticities of the winding.

\item{Cyclic Bias:} Spikes at higher lags would reveal a failure to capture the quasi-periodic nature of higher-order winding numbers ($W=1…4$), indicating that the model is underfitting the global topological context.
\end{itemize}

In our analysis, the transition of the NARX model toward a perfect Kronecker-delta-like ACF serves as further evidence for the discovery of a deterministic identity between distinct topological sectors.

\begin{figure}[h]
\includegraphics[width=1\linewidth]{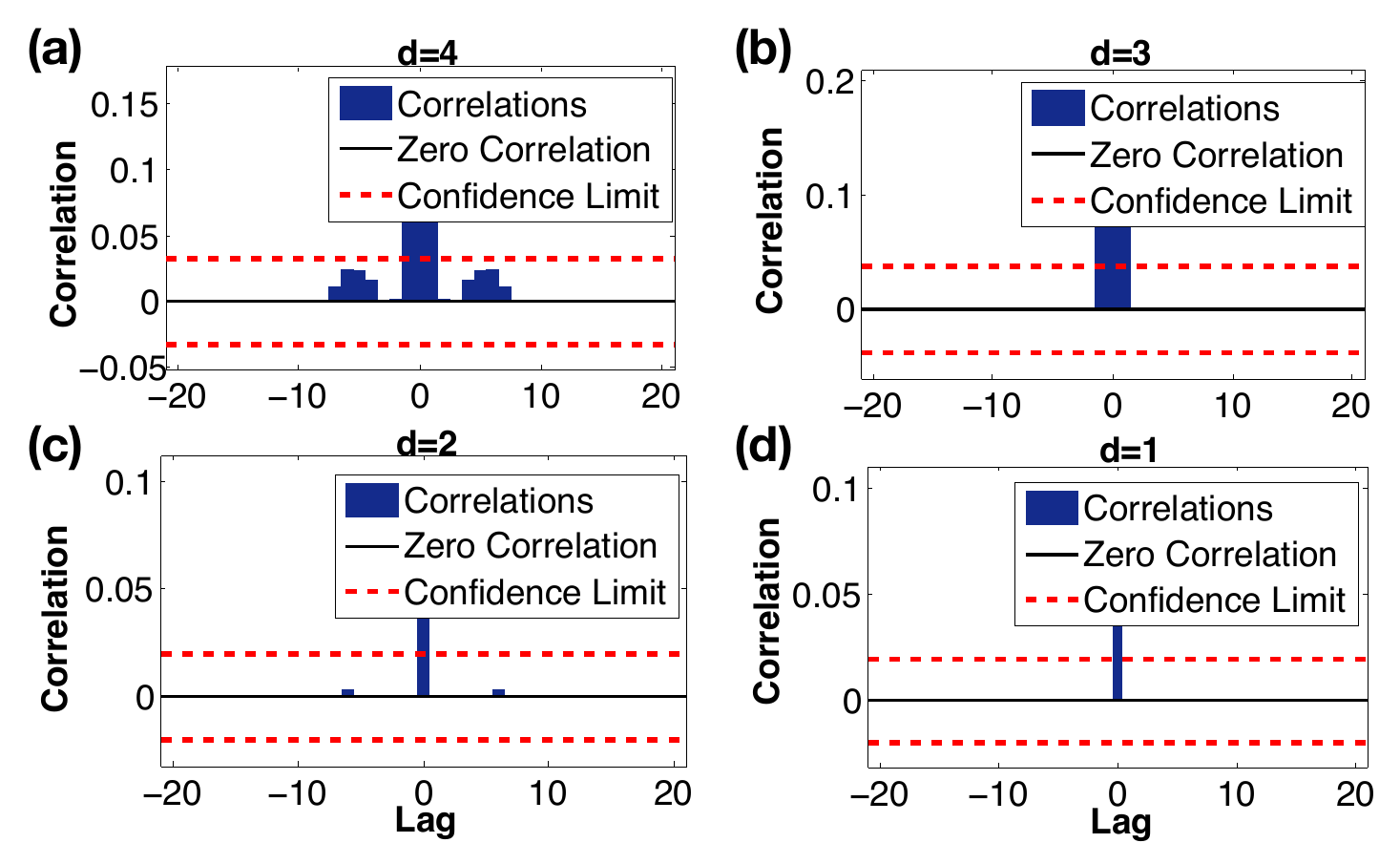}
\caption{\label{fig:frog} NARX neural network auto-correlation error vs. Lag for: (a) $d=4$ past values, (b) $d=3$ past values, (c) $d=2$ past values and (d) $d=1$ past values. Red dashed lines indicate confidence intervals.  }
\end{figure}

\begin{figure}[h]
\centering
\includegraphics[width=1\linewidth]{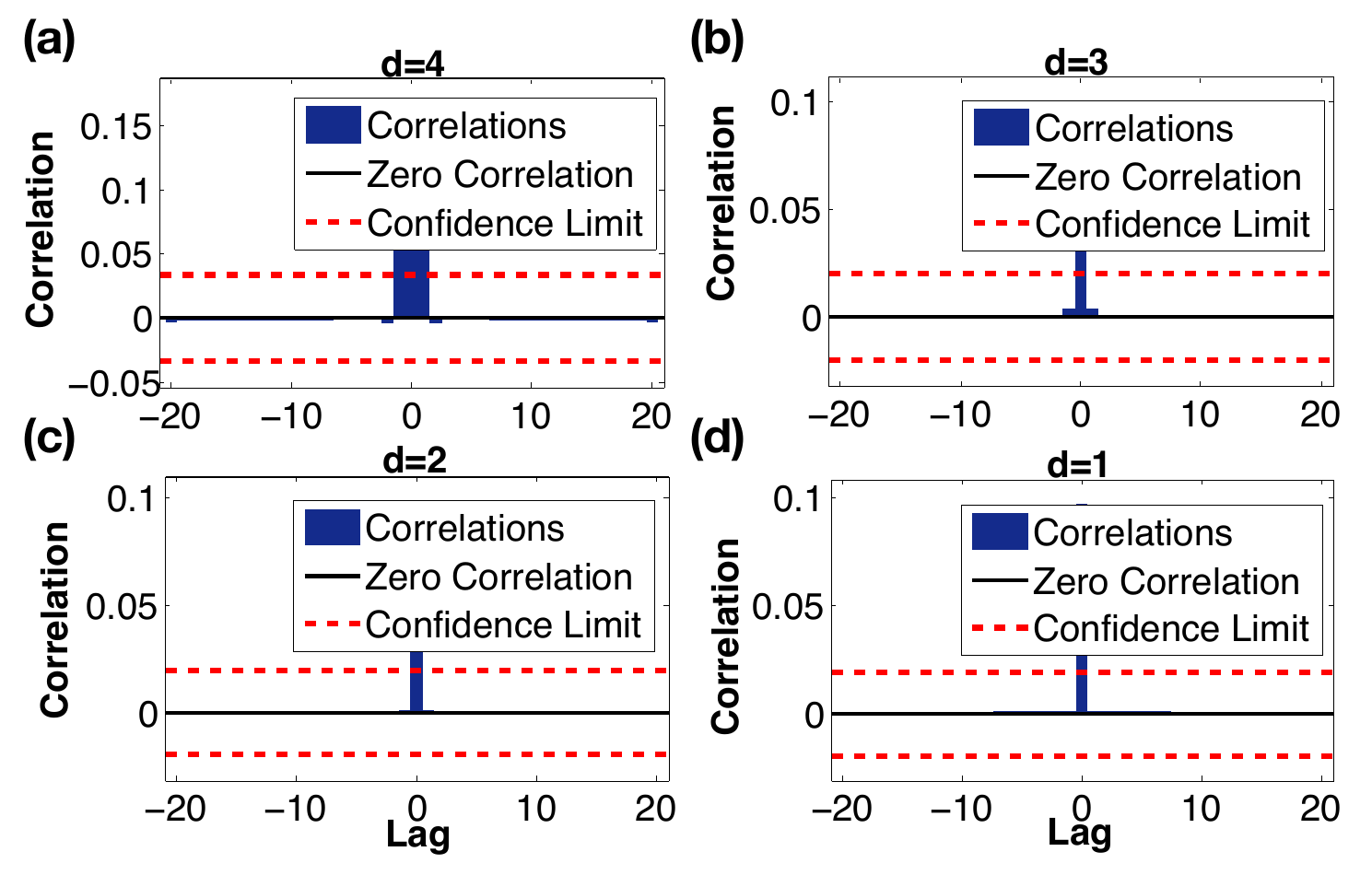}
\caption{\label{fig:frog} NAR neural network auto-correlation error vs. Lag for: (a) $d=4$ past values, (b) $d=3$ past values, (c) $d=2$ past values and (d) $d=1$ past values. Red dashed lines indicate confidence intervals.  }
\end{figure}

\begin{figure}[b]
\includegraphics[width=1\linewidth]{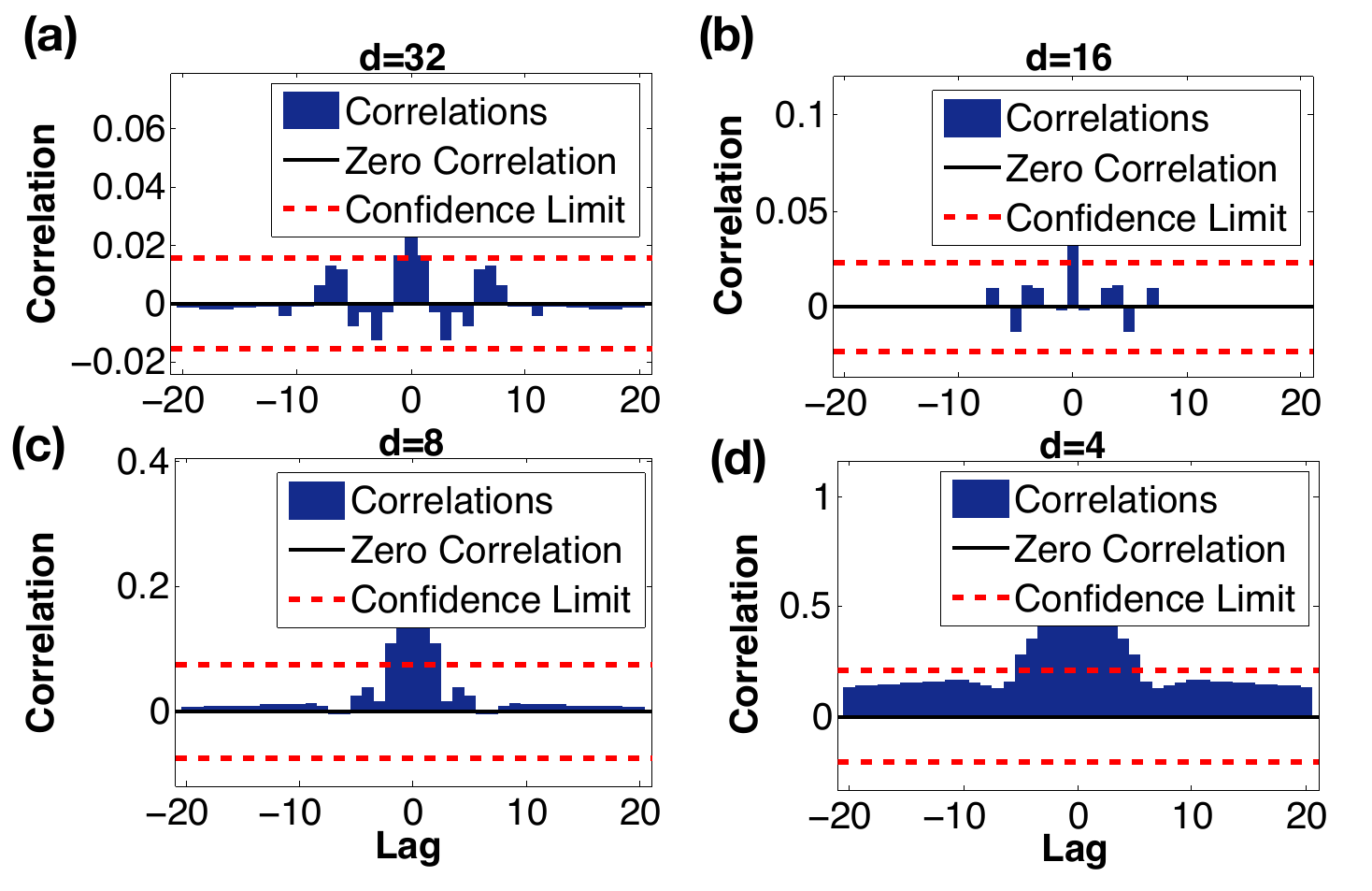}
\caption{\label{fig:frog} NIO neural network auto-correlation error vs. Lag for: (a) $d=32$ past values, (b) $d=16$ past values, (c) $d=8$ past values and (d) $d=4$ past values. Red dashed lines indicate confidence intervals. }
\end{figure}

\subsection{Convergence Dynamics and Performance Limits}

The optimization of the network’s synaptic weights is quantified by the evolution of the Mean Squared Error (MSE) across successive training epochs. We define the performance function as the quadratic loss over the target manifold:
\begin{equation}
MSE= \frac{1}{N} \sum_{i=1}^{N}[X_{\mathrm{output}}(c_{i})-X_{\mathrm{target}}(c_{i})]^2.
\end{equation}

The MSE trajectory provides a diagnostic of the network’s \textquotedblleft learning" of the topological invariants. In an ideal training regime, the MSE exhibits a monotonic decay, representing the systematic reduction of the informational entropy between the model’s prediction and the true quantum state. Below we outline the MSE convergence regimes and precision:

\begin{itemize}

\item{Optimization and Saturation:} The rapid descent in MSE signifies the network's acquisition of the primary jump features at $c_{crit}$. The eventual plateau identifies the saturation limit, where the model has internalized the deterministic structure of the winding number sector.

\item{Generalization vs. Overfitting: }To ensure the robustness of the predictive mapping, we implement a multi-fold validation strategy. A divergence between training and validation loss—where the latter increases despite a diminishing training error—indicates the onset of overfitting (i.e., the network is memorizing the sequence noise rather than the underlying physics). We utilize an early-stopping criterion at the point of maximum generalization to preserve the model's predictive integrity.

\item{The Numerical Floor:} A singular result of this study is the attainment of an MSE on the order of $10^ {-27}$ within the NARX framework. This value significantly surpasses the standard double-precision machine epsilon ($\epsilon \approx 10 ^{-16}$), implying that the gradient of the performance function has effectively vanished. This residual numeral value is not the result of the optimization algorithm, but the direct mathematical consequnce of a numerically null gradient, which shows that NARX architecture with $d=1$ has deterministically mapped the phase space with a global exact minimum, providing computational evidence of the existencie of an analytical closed relation between the winding numbers $W$ and the critical paramater $c_{\mathrm{crit}}$.

\end{itemize}

\begin{figure}[h]
\includegraphics[width=1\linewidth]{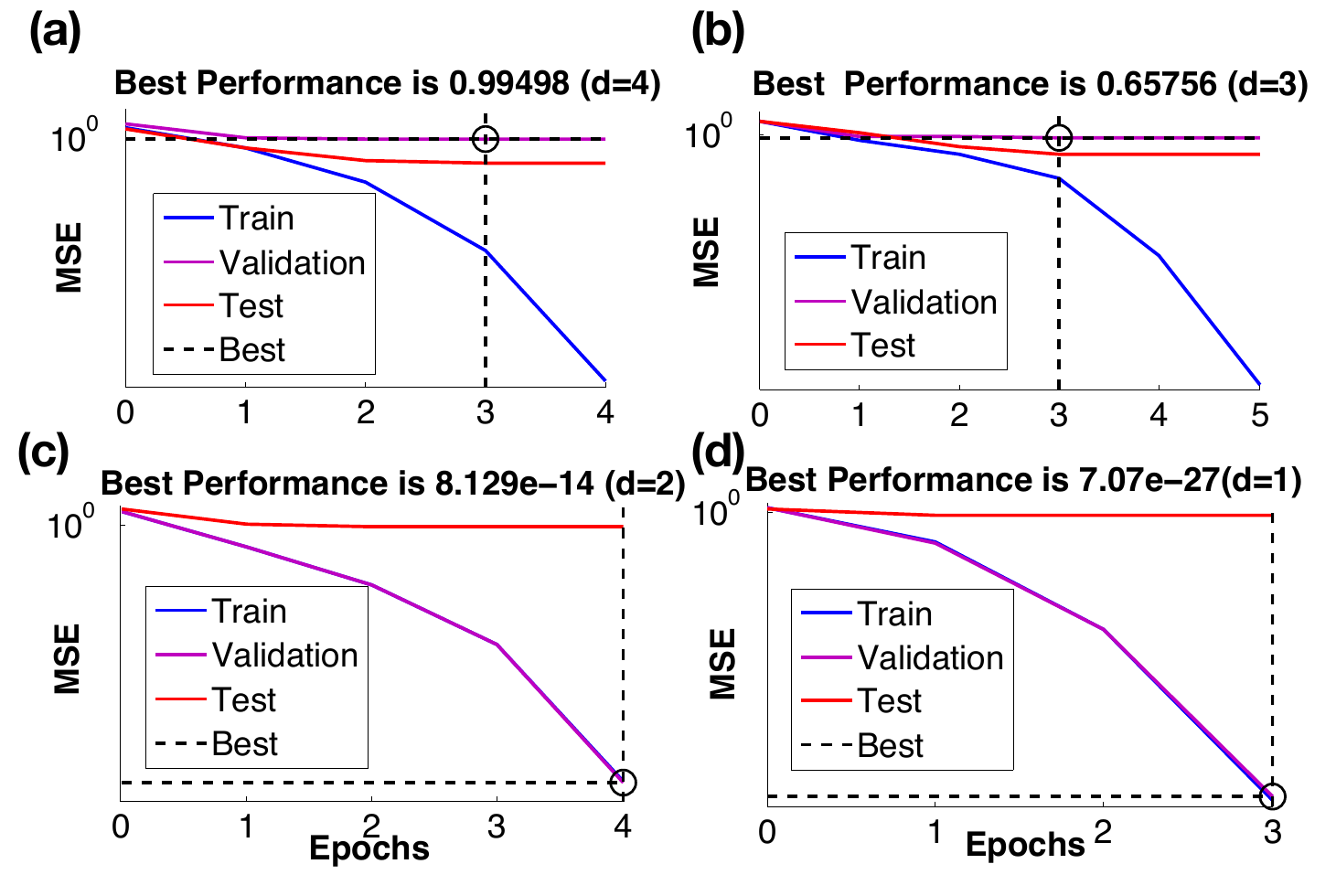}
\caption{\label{fig:frog} NARX neural network performance quantified by the Mean Squared Error (MSE) vs. epochs for:  (a) $d=4$ past values, (b) $d=3$ past values, (c) $d=2$ past values and (d) $d=1$ past values. Best validation performance is indicated with dashed lines. }
\end{figure}

\begin{figure}[h]
\includegraphics[width=1\linewidth]{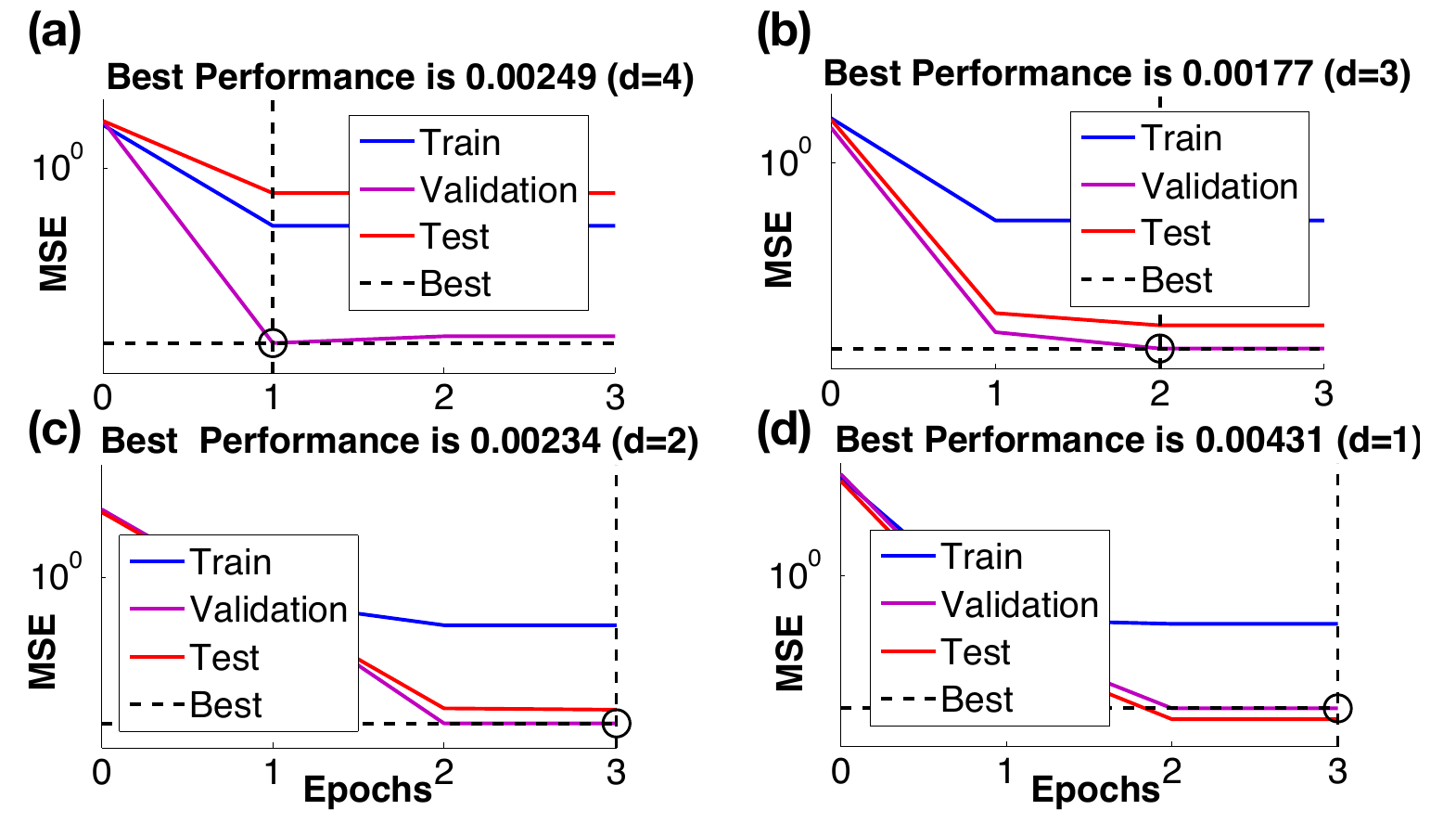}
\caption{\label{fig:frog} NAR neural network performance quantified by the Mean Squared Error (MSE) vs. epochs for:  (a) $d=4$ past values, (b) $d=3$ past values, (c) $d=2$ past values and (d) $d=1$ past values. Best validation performance is indicated with dashed lines. }
\end{figure}

\begin{figure}[b]
\includegraphics[width=1\linewidth]{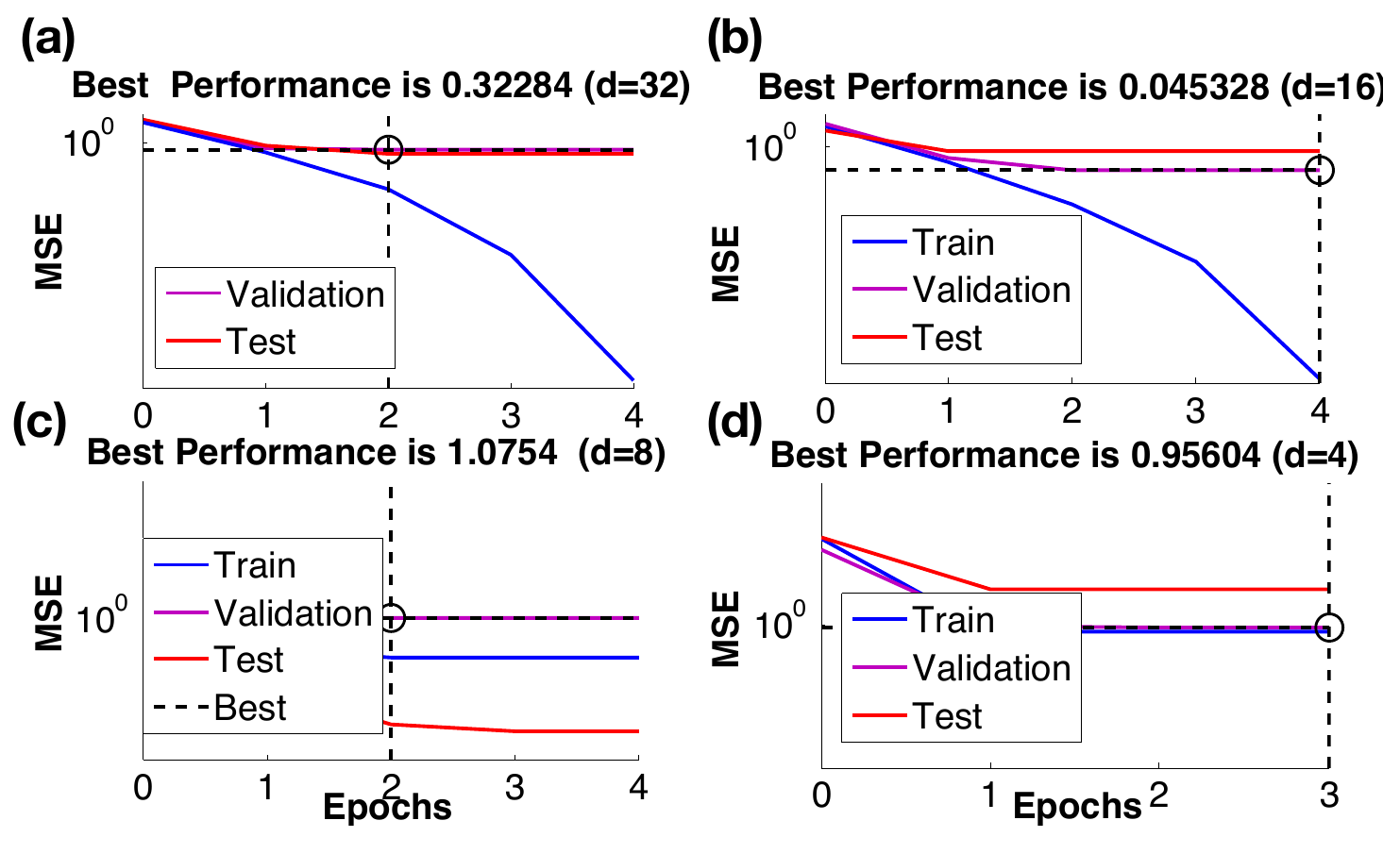}
\caption{\label{fig:frog} NIO neural network performance quantified by the Mean Squared Error (MSE) vs. epochs for:  (a) $d=32$ past values, (b) $d=16$ past values, (c) $d=8$ past values and (d) $d=4$ past values. Best validation performance is indicated with dashed lines. }
\end{figure}

\subsection{Comparative Performance}

The predictive efficacy of the three architectures—NARX, NAR, and NIO—is evaluated through a systematic sweep of the embedding dimension (delay depth $d$). For the NARX and NAR models, we utilize a hidden layer of 10 neurons. In contrast, the NIO architecture is expanded to 20 neurons to compensate for its significantly lower predictive capacity. The target manifold $y(t)$ corresponds to the geometric phase $X(c)$ for $W=5$ topological sector, while the exogenous driving signals $x(t)$ for the NARX and NIO models comprise the sectors $W\in \{1,2,3,4\}$.

\subsubsection{Topological Response and Residual Manifolds}

The global fidelity of the neural network responses is illustrated in Figs. 3, 4, and 5. For the NARX architecture (Fig. 3), we observe a phase-locked convergence to the target sequence. As shown in Figs. 3(a)–(d), the output captures the discrete jump at $c_{crit}=60$ with increasing precision as the delay is optimized. The optimal response is achieved at $d=1$, where the absolute error $\delta(c)$ effectively vanishes. The NAR model (Fig. 4) demonstrates robust local-trend tracking, reaching its highest fidelity at $d=3$. However, unlike the NARX model, it retains a finite residual floor across all tested delays. Conversely, the NIO architecture (Fig. 5) fails to resolve the topological transition reliably. Despite the increased neuronal capacity and a larger search space for delays ($d \in \{4,…,32\}$), the response remains dominated by stochastic noise, confirming that a purely input-output mapping without autoregressive feedback is insufficient for capturing the non-analyticities of the phase transition.

\subsubsection{Residual Whiteness and Autocorrelation}

To verify the informational completeness of the models, we analyze the residual autocorrelation function (ACF) in Figs. 6, 7, and 8. The NARX architecture (Fig. 6) exhibits a transition toward a perfect Kronecker-delta correlation ($R(k) \approx \delta_{k,0}$) for $d=1$. For higher delays ($d=4$), we observe structured residuals that, while remaining within the $95\%$ confidence intervals (dashed lines), indicate that excessive temporal memory introduces systematic interference. In the NAR framework (Fig. 7), the ACF confirms that the residuals are largely decorrelated at $d=3$, yet significant spikes emerge outside the confidence bounds for $d=4$, signaling an unresolved dependency in the self-referential mapping. The NIO ACF (Fig. 8) reveals the most significant divergence; the residuals display non-random, long-range correlations for $d=4$ and $d=8$. A relative stabilization is only achieved at $d=16$, where the network averages over a broader exogenous context to mitigate its lack of internal feedback.

\subsubsection{Convergence to the Numerical Floor}

The learning trajectories and final performance limits are quantified via the Mean Squared Error (MSE) in Figs. 9, 10, and 11. The NAR and NIO models achieve stable convergence at MSE values of $10^{ -3}$ and $10^{ -2}$, respectively. The NIO model achieves its best validation performance (MSE=$0.04$) at $d=16$, although the significant divergence between the training and validation data can be a signature of overfitting. The most striking result is found in the NARX convergence at $d=1$ (Fig. 9 (d)). The NARX architecture achieves an MSE on the order of $10 ^{-27}$ by the third training epoch, effectively reaching the absolute limit of numerical precision for the computational environment. This scale is significantly lower than the standard double-precision machine epsilon, implying that the gradient of the performance function has reached a vanishing point relative to the floating-point representation. This convergence identifies a perfect functional identity, suggesting that the NARX model has successfully resolved an exact analytical mapping between the exogenous winding sectors $W=1,2,3,4$  and the target topological phase transition.\\

Close inspection reveals that the NARX model reaches near-perfect precision at $d=1$, meaning that the relationship between the target winding number ($W=5$) and the previous ones ($W=1,2,3,4$) is instantaneous and  deterministic. When increasing the delay to $d=4$, we are forcing the NN to look at 4 past steps of the exogenous inputs. The fact that the MSE collapses from $10^{-27}$ to $0.99$ suggests that the  context provided by previous winding numbers is only relevant in a very tight, phase-locked window. By considering delays  $d=2,3,4$ we are introducing information that is no longer relevant to the current state, which the network perceives as interference or noise. While NAR remains robust and self-referential across different values of $d$, NARX is acting like a high-magnification microscope: when it is perfectly focused ($d=1$), it sees everything (deterministic identity), but when it is slightly out of focus ($d=4$), the image disappears entirely. Moreover, the results confirm a highly dynamic sensitivity: The fact that the MSE jumps from $10^{-27}$to $0.99$ indicates that the NARX architecture is actively attempting to find a mathematical correlation between the inputs and the target, as opposed to statically reproducing a fixed output.

\subsubsection{Information Gain Analysis}

In order to quantify how much the exogenous input helps the performance of the NARX NN with respect to NAR NN, we calculated the delta metric $\Delta=| MSE_{d=1}^{NARX}-MSE_{d=1}^{NAR}| \approx 0.004$. The NARX model is performing roughly $10^{-24}$  times better than the NAR model. This suggests that the information contained in the previous winding numbers ($W=1,2,3,4$)  is the missing key to the deterministic mapping. This has an interesting interpretation: The NAR model (MSE=0.004) is trying to predict the future of $W=5$ by looking only at its own past. This small error suggests $W=5$ is mostly self-referential but contains some \textquotedblleft unpredictable" complexity or noise. The NARX model (MSE=$10^{-27}$) uses the other $W$ values to eliminate that noise entirely. This means the "noise" in $W=5$ is actually a deterministic signal that is present in $W=1,2,3,4$. 

\subsubsection{Computational Complexity vs. Accuracy}

\begin{table*}[ht]
\centering
\caption{Comparative Analysis of NARX, NAR, and NIO architectures at Fixed Delay $d=4$: Accuracy vs. Network Capacity. Capacity is defined by the number of data points used in the input layer.}
\label{tab:nn_comparison_d4}
\begin{tabular}{@{}lcccc@{}}
\toprule
\textbf{Architecture} & \textbf{MSE ($d=4$)} & \textbf{Input Capacity} & \textbf{Efficiency Rank ($d=4$)} & \textbf{Performance Interpretation} \\ \midrule
\textbf{NAR}          & 0.002                & 100 points              & 1st                      & Local-trend capture         \\
\textbf{NIO}          & 0.956                & 100 points              & 2nd                      & Insufficient state mapping          \\
\textbf{NARX}         & 0.995                & 200 points              & 3rd                      & Signal interference     \\ \bottomrule
\end{tabular}
\end{table*}

\begin{table*}[ht]
\centering
\caption{Comparative performance and sensitivity analysis of NARX, NAR, and NIO architectures for topological phase transition estimation.}
\label{tab:nn_comparison}
\begin{tabular}{@{}lcccc@{}}
\toprule
\textbf{Architecture} & \textbf{Best MSE} & \textbf{Optimal Delay ($d$)} & \textbf{DSI} & \textbf{Memory Strategy} \\ \midrule
\textbf{NARX}         & $10^{-27}$        & 1                            & Highest      & Phase-locked context     \\
\textbf{NAR}          & $0.001$           & 3                            & Lowest       & Local-trend robustness   \\
\textbf{NIO}          & $0.045$           & 16                           & High         & Statistical averaging    \\ \bottomrule
\end{tabular}
\end{table*}

In order to compare computational complexity vs. accuracy, in Table 1 we compare the number of trainable parameters (Capacity) against MSE (Performance) at $d=4$, this is a measure of architectural efficiency. The most striking result here is that NAR performs significantly better than NARX (0.002 vs. 0.99) when the delay is fixed at $d=4$. This means that at this specific delay, the exogenous inputs ($W=1,2,3,4$) are acting as incoherent noise or interference rather than helpful data. This suggests that the relationship between different $W$ is non-local, or highly sensitive to timing. Thus, more data does not imply better results, the perfect mapping requires a specific optimized delay $d=1$ to align the exogenous inputs correctly. In addition, the NIO model has the same number of data points as the NAR model (100) but performs much worse (0.95 vs. 0.002). This implies the system has \textquotedblleft memory." The value of the winding number $W$ at point $c$ is heavily dependent on its value at $c-1$. Because NIO lacks this feedback loop, it is trying to map $x \rightarrow y$ blindly, failing to capture the sequential evolution of the phase transition.

Crucially, while both the NARX and NIO architectures exhibit a comparable macroscopic degradation in performance at delay $d=4$ —as evidenced by their similarly elevated MSE values in Table I—a deeper diagnostic is revealed through close inspection of the Autocorrelation Function (ACF) of the residuals. The marginal distributions of the ACF for the NIO model are significantly worse than those of the NARX counterpart. While the NARX model, even under sub-optimal delay constraints, retains a degree of structural correlation in its residuals due to its inherent autoregressive feedback, the NIO architecture completely decouples from the sequential dependencies of the topological phases. This distinction demonstrates that the failure mode of NARX is a structured consequence of the \textquotedblleft complexity paradox" (i.e., sensitivity to non-local parameter history), whereas the NIO failure represents a fundamental inability of the purely input-driven architecture to track the underlying manifold topology without recursive memory.

\subsubsection{Delay Sensitivity Index (DSI)}

In order to further quantify the numerical results, we calculated the Delay Sensitivity Index (DSI) $DSI=\frac{MSE_{max}-MSE_{min}}{\Delta d}$ for NARX, NAR, and NIO obtaining $DSI=0.33,  0.0015, 0.129$. This is depicted in Table 2. Some insights regarding Table 2: The NAR model only needs $d=3$ to achieve 0.001 accuracy. This suggests that the signal $X$ vs. $c$ is highly self-correlated; the \textquotedblleft memory" of the phase transition is stored locally in its own history. The NIO model’s need for a much larger window ($d=16$) to reach even 0.04 accuracy, suggesting that external winding numbers ($W=1…4$) do not map to $W=5$ directly. They provide a context, but without autoregressive feedback, the network must average over a much longer history to find the state. The NARX model bridges the gap. By combining the high-quality local memory of NAR with the global context of the other winding numbers, it collapses the error to the limit of numerical precision at the absolute minimum delay ($d=1$).

\section{Discussion}

The comparative analysis of the NARX, NAR, and NIO architectures establishes a performance hierarchy that provides a numerical existence proof for a deterministic relationship between winding numbers and critical-parameter transitions. The NARX model emerges as an exact functional estimator, achieving a near-perfect functional identity at minimal delay ($d=1$) and effectively reaching the boundaries of numerical precision to confirm that the underlying physics is mathematically closed. While the NAR model demonstrates that the system is largely self-referential, its higher residual error confirms that exogenous inputs from other winding numbers constitute the missing key to a total deterministic mapping. Conversely, the failure of the NIO architecture---despite its increased neuronal capacity---underscores that topological phase transitions cannot be captured by blind input-output mappings; rather, they require both autoregressive feedback and immediate exogenous context to resolve the exact governing laws of the system.

\subsection{The Hierarchy of Predictive Capacity}
The most striking result is the near-perfect convergence of the NARX architecture. By achieving a Mean Squared Error (MSE) on the order of $10^{-27}$ at $d=1$, the model effectively reaches the limit of numerical precision. This collapses the distinction between an estimate and a perfect functional identity, implying that the critical-measurement-strength parameter $c_{\mathrm{crit}}$ for a given winding number $W$ is mathematically locked to the states of its predecessors ($W=1 \dots 4$). 

In contrast, the NAR model---while reliable---maintains a residual error ($\Delta \approx 0.004$). This gap represents the information gain provided by exogenous inputs. While NAR is largely self-referential, the NARX model's ability to eliminate residual noise entirely suggests that the seemingly unpredictable components of a single topological sequence are actually deterministic signals embedded in the global configuration of the other winding numbers. Globally, this implies that the sequential transitions are not independent events; instead, they are governed by an overarching mathematical law shared across the entire parameter-space trajectory.

\subsection{Phase Sensitivity and the Complexity Paradox}
The analysis of the Delay Sensitivity Index (DSI) and the ``complexity paradox'' at $d=4$ provides a vital validation of the model's integrity. While the NAR architecture proves robust and self-referential, maintaining stable performance across various delays, the NARX model behaves like a high-magnification microscope: for optimal alignment ($d=1$), the network achieves a phase-locked focus, resolving the deterministic identity of the transition. 

This sensitivity confirms that the NARX network is not merely memorizing sequences or producing a trivial static output. Instead, it is performing a high-precision dynamic mapping. The performance collapse at $d=4$ proves that the underlying physical mapping is strictly local within the parameter space. Forcing the network to integrate non-local, asynchronous ``ghosts'' of past states ($t-4$) introduces artificial phase delays that act as destructive interference, disrupting the tight, localized tracking required to resolve the sharp transition.

\subsection{Memory vs. Blind Mapping}
Finally, the failure of the NIO architecture---despite its higher neuron count---underscores the necessity of autoregressive feedback in physical systems with sequential dependence. Without the ability to reference its own history ($y(t-1)$), the NIO model is forced to rely on statistical averaging over much larger windows ($d=16$) to estimate the state, yet it still fails to capture the sharp transition at $c_{\mathrm{crit}}$. 

A close inspection of the Autocorrelation Function (ACF) of the residuals highlights this structural divergence: while both NARX and NIO suffer a macroscopic collapse in MSE at sub-optimal delays ($d=4$), the marginal distributions of the ACF are significantly worse for the NIO architecture. Even under sub-optimal delay constraints, NARX retains a degree of correlation structure in its residuals due to its recursive memory loop. Conversely, the NIO architecture completely decouples from the sequential dependencies of the topological phases, demonstrating that a purely input-driven network cannot track the underlying manifold progression without internal feedback.

\subsection{Underlying Quantum Mechanism}
Beyond the computational efficiency of the autoregressive framework, the exact functional mapping achieved by the NARX model ($d=1$) points to a profound underlying quantum mechanism. In this weak-measurement scheme, tuning the measurement-strength parameter $c$ does not alter the geometry of the spatial manifold---which remains strictly constrained to a fixed equatorial circle on the Bloch sphere---but rather modulates its angular velocity. 

While the geometric phase $X$ is formally evaluated over a closed path where the initial and final states are equal ($|\langle\psi_{\text{final}}|\psi_{\text{initial}}\rangle| = 1$), its value shifts via a sudden, non-analytic topological jump at the critical boundary $c_{\mathrm{crit}}$. Because moving across different topological regions (discretely labeled by winding numbers $W=1 \dots 4$) systematically alters how fast the quantum state winds along the equator, the precise location of the topological jump for a higher sector (e.g., $W=5$) is structurally predetermined by the historical velocity profile. The exceptional capacity of the NARX network to resolve this boundary at $d=1$ implies that it successfully acts as a high-precision numerical probe, mapping the continuous, step-by-step acceleration of the state vector along the equator before it hits the phase discontinuity.

\subsection{Outlook and Future Frameworks}

While the current NARX framework demonstrates remarkable precision, future iterations could transition from purely predictive modeling to more complex classification tasks. To evaluate the scalability of the learned physical laws, a key next step involves testing generalization capabilities across higher winding numbers (e.g., training on $W=1\dots4$ to predict $W \ge 10$). Additionally, introducing synthetic noise into the input sequences will be essential to simulate realistic experimental environments and evaluate the stability of the $c_{\mathrm{crit}}$ estimation. 

Furthermore, exploring Recurrent Neural Networks (RNNs) or Long Short-Term Memory (LSTM) units could offer advanced handling of the temporal dependencies inherent in the $X$ vs. $c$ sequences, potentially mitigating the limitations observed in the NIO model. Finally, systematic ablation studies---where $W$ sequences are removed sequentially---will allow for the isolation of specific exogenous contributions, explicitly revealing the correlation distance between distinct topological sectors.


\section{Conclusion}

In summary, the comparative analysis of the NARX, NAR and NIO dynamic neural network architectures reveals that while the NAR model efficiently captures local trends through self-referential \textquotedblleft memory," and the NIO model fails by attempting a blind statistical mapping without feedback, the NARX architecture achieves a state of deterministic identity. By collapsing the Mean Squared Error (MSE) to the limit of numerical precision (MSE=$10^ {-27}$) at a phase-locked delay of $d=1$, NARX demonstrates that the \textquotedblleft noise" inherent in a single winding number sequence is actually a deterministic signal provided by the context of exogenous topological sectors. The dramatic performance collapse of NARX at $d=4$ further validates this finding, serving as a \textquotedblleft complexity paradox" that proves the model is not merely memorizing data but is instead acting as a high-precision instrument sensitive to the exact temporal alignment of the phase transition. Consequently, this study confirms that the relationship between winding numbers $W$ and critical-measurement-strength parameter $c_{\mathrm{crit}}$ is mathematically deterministic, and can be unlocked provided the neural network architecture accounts for both autoregressive history and immediate exogenous context.

\begin{acknowledgments}
    
The author is grateful to Yuval Gefen, Kyrylo Snizhko, and Alessandro Romito for significant assistance in the core numerical codes developed for the geometric phase via weak measurements framework. The generation of the topological sequences for different winding numbers $W$, as well as the neural network implementation, was conducted independently by the author. The author acknowledges financial support from FONCyT (PICT Startup 2015 0710). 
\end{acknowledgments}

\end{document}